\def\dend#1{{\if*#1{\it Paenibacillus dendritiformis}\else
                {\it P. dendritiformis}\fi}}
\def\Tvar{var. {\it dendron}}
\def\Cvar{var. {\it chiralis}}
\def\Tname#1{{\if*#1\dend* \Tvar\else
                \if-#1\dend{} \Tvar\else
                 \dend{} \Tvar{} #1\fi\fi}}
\def\Cname#1{{\if*#1\dend* \Cvar\else
                \if-#1\dend{} \Cvar\else
                 \dend{} \Cvar{} #1\fi\fi}}
\def\Vname#1{{\if*#1\eddi{Paenibacillus}{V}\else
                \if-#1\eddi{P.}{V}\else \eddi{P.}{V} #1\fi\fi}}
\def\bsub#1{{\if*#1{\it Bacillus subtilis}\else
                \if-#1{\it B. subtilis}\else {\it B. subtilis} #1\fi\fi}}
\def\bacil#1{\if *#1{Bacillus}\else{B.}\fi}
\def\bcirc#1{\if *#1{\it \bacil* circulans}\else
                \if -#1{\it \bacil{} circulans}\else
                   {\it \bacil{} circulans} #1\fi\fi}
\def\ecoli#1{{\if*#1{\it Escherichia coli}\else
                \if-#1{\it E. coli}\else {\it E. coli} #1\fi\fi}}
\def\salmon#1{{\if*#1{\it Salmonella typhimurium}\else
                \if-#1{\it S. typhimurium}\else
                {\it S. typhimurium} #1\fi\fi}}

\def\myxo#1{{\if*#1{\it Myxococcus xanthus}\else
                \if-#1{\it M. xanthus}\else
                {\it M. xanthus} #1\fi\fi}}

\def\be{\begin{equation}}
\def\ee{\end{equation}}
\def\ben{\begin{enumerate}}
\def\een{\end{enumerate}}
\def\ba{\begin{eqnarray}}
\def\ea{\end{eqnarray}}
\def\partderiv#1#2{{\partial #1\over\partial #2}}

\def\etal{{\it et al. }}

\def\text#1{\hbox{#1}}

\documentstyle[fleqn,pre,aps,preprint,epsf]{revtex}

\begin{document}
\draft

\title{Spatio-selection in Expanding Bacterial Colonies}
\author{Ido Golding, Inon Cohen and Eshel Ben-Jacob}
\address{School of Physics and Astronomy, Raymond and Beverly Sackler
  Faculty of Exact Sciences, \\ Tel Aviv University,
Tel Aviv 69 978, Israel}
\maketitle


\begin{abstract}

Segregation of populations is a key question in evolution theory. One 
important aspect is the relation between spatial organization 
and the population's composition. Here we study a specific example 
-- sectors in expanding bacterial colonies. Such sectors are 
spatially segregated sub-populations of mutants. The sectors can be 
seen both in disk-shaped colonies and in branching colonies. We study 
the sectors using two models we have used in the past to study 
bacterial colonies -- a continuous reaction-diffusion model  with 
non-linear diffusion and a discrete ``Communicating Walkers'' model. We find that in 
expanding colonies, and especially in branching colonies, segregation 
processes are more likely than in a spatially static population. 
One such process is the establishment of stable sub-
population having neutral mutation. Another example is the maintenance
of wild-type population along side with sub-population of advantageous
mutants. Understanding such processes in bacterial colonies is an
important subject by itself, as well as a model system
for similar processes in other spreading populations.

\end{abstract}



\section{Introduction}
\label{sec:experiments}

Charles Darwin's theory of evolution was inspired by his observations
at the Galapagos Islands \cite{Darwin1859}.
These observations of small 
differences between related species led him to identify the importance 
of mutations and to create the theory of natural selection. The 
speciation he saw was intimately related to spatial structure -- the 
geographical separation of the islands -- and to temporal dynamics -- 
the spreading of birds and animals from island to island. The 
geographical separation prevented the occupants of a new island from 
being mixed back into the population from which they came. The 
geographical separation soon turned into segregation of the population 
into genetically different groups -- the first step towards speciation. 
Aside from natural selection, another evolutionary "force" acting in this 
case is genetic drift. Genetic drift is the process where due to 
fluctuations in a small population a mutation can spread in the entire 
population even if it is selectively neutral or even slightly 
disadvantageous.

Recently, pattern formation in bacterial colonies became the focus of 
attention 
\cite{BL98,MKNIHY92,MS96,Kessler85,FM89,PK92a,BSST92,MHM93,BSTCCV94a,BCSALT95,WTMMBB95,BCCVG97,KW97,ES98}.
In this paper we study the subject of expression of 
mutations, and especially segregation of populations, in the context of 
expanding bacterial colonies. We take the bacterial colonies to be a 
model system for the study of expression of mutations in a spreading 
population, as well as an interesting and important subject by itself. 

During the course of evolution, bacteria have developed sophisticated 
cooperative behavior and intricate communication capabilities 
\cite{SDA57,Shap88,BenJacob97,BCL98,LB98}. These include: 
direct cell-cell physical interactions via extra- membrane polymers 
\cite{Mend78,Devreotes89}, collective production of extracellular 
"wetting" fluid for movement on hard surfaces 
\cite{MKNIHY92,Harshey94}, long range chemical signaling, such as 
quorum sensing \cite{FWG94,LWFBSLW95,FWG96} and 
chemotactic signaling\footnote{Chemotaxis is a bias of movement 
according to the gradient of a chemical agent. Chemotactic signaling is 
a chemotactic response to an agent emitted by the bacteria.} 
\cite{BB91,BE95,BB95}, collective activation and deactivation of 
genes \cite{ST91,SM93,MS96} and even exchange of genetic material 
\cite{GR95,RPF95,Miller98}. Utilizing these capabilities, bacterial 
colonies develop complex spatio-temporal patterns in response to 
adverse growth conditions.

Fujikawa and Matsushita \cite{FM89,MF90,FM91} reported for the 
first time\footnote{We refer to the first time that branching growth 
was studied as such. Observations of branching colonies occurred long 
ago \protect\cite{SC38,Henrici48}. } that bacterial colonies could 
grow elaborate branching patterns of the type known from the study of 
fractal formation in the process of diffusion-limited-aggregation (DLA) 
\cite{WitSan81,Sander86,Vicsek89}. This work was done with \bsub*, 
but was subsequently extended to other bacterial species such as {\it 
Serratia marcescens} and {\it Salmonella anatum} \cite{MM95}. It 
was shown explicitly that nutrient diffusion was the relevant dynamics 
responsible for the growth instability. 
Motivated by these observations, Ben-Jacob \etal 
\cite{BSST92,BTSA94,BSTCCV94a} conducted new experiments to 
see how adaptive bacterial colonies could be in the presence of external 
"pressure", here in the form of a limited nutrient supply and hard 
surface. The work was done with a newly identified species, \Tname* 
\cite{TBG98}. This species is motile on the hard surface and its 
colonies exhibit branching patterns (Fig. \ref{fig:physa1}). 

There is a well known observed (but rarely studied) phenomenon of 
bursts of new sectors of mutants during the growth of bacterial 
colonies (see for example Fig. \ref{fig:shapiro} and Refs
\cite{ST91,Shapiro95b}). 
Actually, the phenomenon is more general. Fig. \ref{fig:boschke}
(taken from \cite{BB98}) shows
an emerging sector in a yeast colony.
If the mutants have the same 
growth dynamics as the "normal", wild-type, bacteria they will usually go 
unnoticed (unless some property such as coloring distinguish them)
\footnote{Different coloring may result from different enzymatic activity
(natural coloring) or from a different response to a staining process
(artificial coloring). In both cases the mutation is not neutral in the
strictest sense, but it is neutral as far as the dynamics is concerned.}. If, 
however, the mutants have different growth dynamics, a distinguished 
sector with a different growth pattern might indicate their presence.

In a branching colony, the geometrical structure may aid the 
bursting of a sector of "neutral" mutants; once a branch (or a cluster 
of branches) is detached from his neighboring branches (detached in 
the sense that bacteria cannot cross from one branch to the other), the 
effective population is smaller than the colony's population. In such a 
"reduced" population, genetic drift is more probable and a neutral 
mutant may take over the population in some branches. Sectors of 
"neutral" mutations usually go undetected -- by definition their growth dynamics 
is identical to that of the wild-type (original) bacteria and no 
geometrical feature highlights the sectors.

Sectors of advantageous mutation are much easier to detect, as they 
usually grow in a somewhat different pattern. An advantage in this 
context might be faster multiplication, higher motility or elevated 
sensitivity to chemotactic materials. In all those cases the mutants have 
an advantage in accessing food resources. In a pre-set geometry (or 
without spatial structure) the mutants might starve the wild-type 
bacteria and drive them to extinction. But in a spreading colony each 
part of the colony is heading in a different direction, thus the two 
populations can co-exist. The dynamic process of spreading aids the 
segregation of the population.



The first analytical study of spatial spread of mutations was done by 
Fisher \cite{Fisher37}. He studied the spread of advantageous 
mutation in the limit of large, spatially uniform population, using the 
Fisher-Kolmogorove equation. This equation describes the time 
evolution of a field representing the fraction of the mutants in the local 
population. The same equation can be taken to describe the spreading 
of a population into an uninhabited space, in which case the field 
represents the density of the bacteria. To study mutants by this 
description one must extend the model to include two fields standing 
for two different types of bacteria. Since these equations are expressed 
in the continuous limit, it excludes a-priori the effect of genetic drift. 
As we discuss elsewhere \cite{GKCB98}, the Fisher equation has 
other shortcomings that make it unsuitable for modeling bacterial 
colonies.



To study the sectors in the bacterial colonies we use generic models, 
i.e. models that adhere as much as possible to biological data, but only 
to details which are needed to understand the subject. The generic 
models can be grouped into two main categories:
1. Continuous or reaction-diffusion models \cite{PS78,Mackay78}. In 
these models the bacteria are represented by their 2D density, and a 
reaction-diffusion equation of this density describes their time 
evolution. This equation is coupled to the other reaction-diffusion 
equations for fields of chemicals, such as nutrient.
2. Discrete models such as the Communicating Walkers model of Ben-
Jacob \etal \cite{BSTCCV94a,BCSCV95,BCCVG97} and the Bions 
model of Kessler and Levine \cite{KL93,KLT97}. In this approach, 
the microorganisms (bacteria in the first model and amoebae in second) 
are represented by discrete entities (walkers and bions, respectively) 
which can consume nutrient, reproduce, perform random or biased 
movement, and produce or respond to chemicals. The time evolution 
of the nutrient and the chemicals is described by reaction-diffusion 
equations. In the context of branching growth of bacterial colonies, the 
continuous modeling approach has been pursued recently by Mimura 
and Matsushita \etal \cite{Mimura97,MWIRMSM98}, Kawasaki \etal 
\cite{KMMUS97} and Kitsunezaki \cite{Kitsunezaki97}. In 
\cite{GKCB98} we present a summary and critique of this approach 
(also see \cite{Rafols98}).

In the current study, we use both discrete and continuous models, 
altered to include two bacterial types. In some cases (but not all) the 
two bacterial types have different growth dynamics. We begin each run 
with a uniform population. The event of mutation is included with 
some finite probability of the wild-type strain changing into a mutant 
during the process of multiplication.

Representing mutations in the above two modeling schemes gives rise 
to possible problems.
In a continuous model there is a problem representing a single 
mutation because the equations deal with bacterial area density, not 
with individual bacterium. In a previous paper we have studied 
\cite{CGKB98} the inclusion of finite size effects in the continuous 
model via a cutoff in the dynamics. For the study of mutations, we use 
as our basic "mutation unit" the cutoff density (see below). The value 
of this density is in the order of a single bacterium in an area 
determined by the relevant diffusion length (the idea of using a cutoff 
density to represent discrete entities was first raised by Kessler and 
Levine \cite{KL98}). 
In the discrete model, each "walker" represents not one bacterium, but 
many bacteria ($10^4 - 10^6$) \cite{BSTCCV94a}. Thus, a mutation 
of one walker means the collective mutation of all the bacteria it 
represents.

Note that in this paper we do not discuss the origin of the mutations. 
The common view in biology is that all mutations are random. Both 
Darwin's original view \cite{Darwin1859a} and modern experiments in 
microbiology \cite{COM88,Hall88,Hall91,Foster93} suggest the 
possibility of mutations designed by the bacteria as a response to a 
specific stressful condition. Since the stress in this case cannot be
accurately assessed by a single bacterium, another possibility is that the 
colony as a whole designs the mutation in response to the 
environmental conditions. It is not necessary that only the descendants 
of a single cell will have the mutation; the bacteria have the means 
\cite{GR95,RPF95,Miller98} to perform "horizontal inheritance" i.e. to transfer 
genetic information from cell to cell. If such autocatalytic or 
cooperative mutation occurs in the experiments, then a mutating 
walker in the Communicating Walkers model might be an accurate 
model after all.

In the next section we present the experimental observations of bacterial 
colonies, mutations and sectors in bacterial colonies. In section 
\ref{sec:models} we present the two models used in this study. Section 
\ref{sec:results} presents the study itself; the simulated colonies, the 
results, and comparison with the experimental observation. We conclude 
(section \ref{sec:discussion}) with a short discussion of the results and 
possible implications to other issues, such as growth of tumors and 
diversification of populations.

\section{Observations of Colonial Development}

In this section, we focus on the phenomena observed during the growth
of \Tname.

\subsection{Growth Features}
As was mentioned, the typical growth pattern on semi-solid agar is a branching pattern,
as shown in Fig. \ref{fig:physa1}. The structure of the branching pattern
varies for different growth conditions, as is demonstrated in
Fig. \ref{fig:physa2}.

Under the microscope, bacterial cells are seen to perform a random-walk-like
movement in a layer of fluid on the agar surface (Fig. \ref{fig:finger}).
This wetting fluid is assumed to be excreted by the
cells and/or drawn by the cells from the agar \cite{BSTCCV94a,BSTCCV94b}.
The cellular movement is confined to this fluid; isolated cells spotted on
the agar surface do not move. 
The fluid's boundary thus defines a local
boundary for the branch. Whenever the cells are active, the boundary
propagates slowly, as a result of the cellular movement and production of
wetting fluid.

The various colonial patterns can be grouped into several ``essential
patterns'' or morphologies \cite{BSTCCV94a,BTSA94}.
In order to explain the various growth morphologies, we have suggested
that bacteria use {\em chemotactic signaling} when confronted with
adverse growth conditions \cite{BSTCCV94a,CCB96,BSCTCV95}.
Chemotaxis means changes in the movement of the cell in response to 
a gradient of certain chemical fields 
\cite{Adler69,BP77,Lacki81,Berg93}. The movement is biased along 
the gradient either in the gradient direction (attractive chemotaxis 
towards, for example, food) or in the opposite direction (repulsive 
chemotaxis away from, for example, oxidative toxins). 
Usually chemotactic response means a response to an externally 
produced field as in the case of chemotaxis towards food. However, 
the chemotactic response can be also to a field produced directly or 
indirectly by the bacterial cells. We will refer to this case as 
chemotactic signaling. 

At very low agar concentrations, $0.5 \% $ and below, the colonies
exhibit compact patterns, as shown in Fig. \ref{fig:sec11}. Microscopic
observations reveal that in this case, the bacteria swim within the
agar. Thus, there is no ``envelope'' to the colony, and hence no
branching pattern emerges (see \cite{GKCB98,KCGB98}).

\subsection{Observations of Sectors}
The bursting of sectors can be observed both during compact and
branching growth. 
Examples of the first kind are shown in
Figs. \ref{fig:sec346}, while examples of
the latter are shown in
Figs.
\ref{fig:sectors12},\ref{fig:sec57}.
As we can see from the pictures, sectors emerging during branching
growth have a greater variety of structure and shapes than those
emerging from compact colonies..
This is demonstrated by Fig. \ref{fig:sectors12} depicting colonies at intermediate levels of
nutrients and agar, and Fig. \ref{fig:sec57}, showing colonies
grown at high nutrient level and at the presence of antibiotics. Note
that the sector on the left side of Fig. \ref{fig:sec57} is much more expanded than that
on the right, probably because it has irrupted at an earlier
stage of the colonial development \footnote{Such an early irrupted sector
  might indicate a mixed population in the initial inoculum, and not a new mutant.}.

Throughout this paper, we use the term ``mutant'' to describe the
population in the emerging sectors. We have not verified the existence
of a genetic difference between the bacteria in the sector and those
in the rest of the colony. We have, however, verified that the
phenotypic difference between the two populations is inheritable,
using inoculation. 

Below we shall see that it is
sometimes possible to relate the shape of the sector, and the
way it ``bursts'' out of the colony, with the specific kind of advantage
that the mutants possess over the original bacteria.

\section{Models}
\label{sec:models}
We now describe the two models we have used to study the development
of a colony consisting of two bacterial strains.
Both models are based on the ones originally used to study a single
strain colony \cite{GKCB98,BCL98,CGKB98,BSTCCV94a}. 

\subsection{The Continuous Model}
A number of continuous models have been
proposed to describe colonial development \cite{KL98,Kitsunezaki97,MWIRMSM98,KMMUS97}.
Following \cite{GKCB98}, the model we take includes a linear growth
term and a non-linear diffusion of the bacteria (representing the
effect of a lubricating fluid \cite{KCGB98}). In the case of a single strain, the
time evolution of the 
2D bacterial density $b({\bf x},t)$ is given by:
\begin{eqnarray}
\label{onespec-eqn}
\lefteqn{ \partderiv{b}{t} = \nabla (D(b) \nabla
  b) + \varepsilon n b \Theta (b-\beta) - \mu b}
\end{eqnarray}
The first term on the RHS describes the bacterial movement,
with $D(b)=D_0 b^k$ ($D_0$ and $k>0$ constants) \cite{Kitsunezaki97}.
The second term is the population growth, which is proportional to
food consumption. 
$n({\bf x},t)$ is the nutrient 2D density and
$\varepsilon$ the nutrient$\rightarrow$bacteria
conversion factor.
The growth term is multiplied by a step function $\Theta(b-\beta)$,
which sets it to zero if the bacterial density is smaller then a
threshold
$\beta$. This threshold represents the discreteness of the bacteria
\cite{KL98}.
We have previously shown that the effect of the step function is
negligible for small $\beta$ \cite{CGKB98}, but we also use it when
implementing the modeling of mutations.
The third term describes bacterial transformation into stationary, pre-spore state, with  $\mu$ the sporulation rate.

In this model, the time development equations for the nutrient concentration $n({\bf x},t)$ and the
stationary bacteria concentration $s({\bf x},t)$ are given by:
 \begin{eqnarray}
\label{onespec-eqn2}
\lefteqn{ \partderiv{n}{t} = D_n\nabla ^2 n - \varepsilon b n \Theta (b-\beta)}\\
\lefteqn{ \partderiv{s}{t} = \mu b}
\end{eqnarray}


We include the effect of chemotaxis in the model using a
{\em chemotactic flux} $\vec{J}_{chem}$, which is written (for
the case of a chemorepellant) as:
\be
{\vec{J}_{chem} = \zeta(b) \chi (r) \nabla r}
\label{j_chem}
\ee
where $r({\bf x},t)$ is the concentration of the chemorepellent agent,
$\zeta(b) = b \cdot b^k = b^{k+1}$ is the bacterial response to
the chemotactic agent \cite{GKCB98},
 and 
$\chi(r)$ is the chemotactic sensitivity to the repellant, which is
negative for a chemorepellant.
In the case of a chemoattractant, e.g. a nutrient, the expression for
the flux will have an opposite sign.
In the case of the ``receptor law'', the sensitivity $\chi(r)$ takes
the form \cite{Murray89}:

\be
{\chi(r) = \frac{\chi_0 K}{(K+r)^2}}
\ee
with $K$ and $\chi_0$ constants.

The equation for $r({\bf x},t)$ is:
\be
{\partderiv{r}{t} = D_r \nabla^2 r + \Gamma_r s - \Omega_r b r - \lambda_r r}
\label{r-eqn}
\ee
where $D_r$ is the diffusion coefficient of the chemorepellent agent,
$\Gamma_r$ is the emission rate of repellant by the pre-spore cells, 
$\Omega_r$ is the decomposition rate of the repellant
by active bacteria,
and $\lambda_r$ is the rate of spontaneous decomposition of the
repellant.

In order to generalize this model to study mutants,
we must introduce two fields, for the densities of the wild-type
bacteria (``type 1'') and the mutants (``type 2''), and allow some
probability of transition from wild-type to mutants.
In the absence of chemotaxis, the equations for the bacterial density
of the two strains will be written
(with subscript denoting bacteria type):
\begin{eqnarray}
\label{twospec-eqn}
\lefteqn{ \partderiv{b_1}{t} = \nabla (D_1(b) \nabla b_1) +
  \varepsilon_1 n b_1 \Theta (b-\beta) - \mu_1 b_1 - F_{12}}\\
\lefteqn{ \partderiv{b_2}{t} = \nabla (D_2(b) \nabla b_2) +
  \varepsilon_2 n b_2 \Theta (b-\beta) - \mu_2 b_2 + F_{12}}
\end{eqnarray}
where $D_{0_{1,2}}=D_{0_{1,2}} b^k$ ($b=b_1+b_2$). 

Note that the mutant strain $b_2$ includes a ``source'' term $F_{12}$,
which is the rate of transition $b_1 \rightarrow b_2$, and is given by
the growth rate of $b_1$ multiplied by a constant mutation rate
(For simplicity, we do not include the process of reverse mutations $F_{21}$. 




\subsection{The Communicating Walkers Model}
\label{subsec:walker}
The Communicating Walkers model \cite{BSTCCV94a} 
is a hybridization of the ``continuous'' and ``atomistic'' 
approaches used in the study of non-living systems. The diffusion of 
the chemicals is handled by solving a continuous diffusion equation 
(including sources and sinks) on a tridiagonal lattice. The bacterial cells are represented by walkers 
allowing a more detailed description. In a typical experiment there 
are $10^9-10^{10}$ cells in a petri-dish at the end of the growth. 
Hence it is impractical to incorporate into the model each and every 
cell. Instead, each of the walkers represents about $10^4-10^5$ cells 
so that we work with $10^4-10^6$ walkers in one numerical 
``experiment''. 

The walkers perform an off-lattice random walk on a plane within an 
envelope representing the boundary of the wetting fluid. This 
envelope is defined on the same triangular lattice where the 
diffusion equations are solved. To incorporate the swimming of the 
bacteria into the model, at each time step each of the active walkers 
(motile and metabolizing, as described below) moves a step of size 
$d$ at a random angle $\Theta $.
If this new position is outside the envelope, the walker does not 
move. A counter on the segment of the envelope which would have been 
crossed by the movement is 
increased by one. When the segment counter reaches a specified number 
of hits $N_c $, the envelope propagates one lattice step and an 
additional lattice cell is added to the colony. This requirement of 
$N_c $ hits represent the colony propagation through wetting of 
unoccupied areas by the bacteria. Note that $ N_c $ is related to the 
agar dryness, as more wetting fluid must be produced (more 
``collisions'' are needed) to push the envelope on a harder 
substrate. 

Motivated by the presence of a maximal growth rate of the bacteria 
even for optimal conditions, each walker in the model consumes food 
at a constant rate $\Omega_c $ if sufficient food is available.
We represent the metabolic state of the $i$-th walker by an 'internal
energy' $E_i $. The rate of change of the internal energy is given by
\begin{equation} 
{\frac{d E_i}{d t }} = \kappa C_{consumed} - {\frac{E_m }{\tau_R}}~, 
\end{equation} 
where $\kappa $ is a conversion factor from food to internal energy 
($\kappa \cong 5\cdot 10^3 cal/g$) and $E_m $ represent the total 
energy loss for all processes over the reproduction time $\tau_R$, 
excluding energy loss for cell division. $C_{consumed}$ is  
$ 
C_{consumed} \equiv \min \left( \Omega_C , \Omega_C^{\prime}\right)~, 
$ 
where $\Omega_C^{\prime}$ is the maximal rate of food consumption as 
limited by the locally available food.
When sufficient food is available, $E_i$ increases until it reaches a 
threshold energy. Upon reaching this threshold, the walker divides 
into two. When a walker is starved for long interval of time, $E_i$ 
drops to zero and the walker ``freezes''. This ``freezing'' 
represents entering a pre-spore state. 

We represent the diffusion of nutrients by solving the diffusion 
equation for a single agent whose concentration is denoted by 
$n(\vec{r},t)$:  
\begin{equation} 
{\frac{\partial n}{\partial t}}=D_n \nabla^2C - b C_{consumed}~, 
\end{equation} 
where the last term includes the consumption of food by the walkers 
($b $ is their density). The equation is solved on the tridiagonal 
lattice. The simulations are started with inoculum of walkers at the 
center and a uniform distribution of the nutrient. 

When modeling chemotaxis performed by walkers, it is possible to
modulate the periods between tumbling (without changing the speed) in
the same way the bacteria do. It can be shown that step length
modulation has the same mean effect as keeping the step length
constant and biasing the direction of the steps (higher probability
to move in the preferred direction). As this later approach is
numerically simpler, this is the one implemented in the Communicating
Walkers model.
As in the aforementioned continuous model, an additional equation is
written for the time evolution of the chemorepellant.

When dealing with two bacterial strains, each walker in the simulation
belongs to either ``type 1'' (wild-type) or ``type 2'' (mutant), which
may differ in their various biological parameters, such as step length
(i.e.  motility) or sensitivity to
chemotaxis. The colony is initialized with an inoculation of wild-type
walkers, which -- when multiplying -- have some finite probability of
giving birth to a mutant. The two populations than co-evolve
according to the dynamics described above.

\section{Results}
\label{sec:results}
\subsection{Compact growth}
We start by examining mutations in colonies grown on soft agar, where
growth is compact. We expect that in
this case a neutral mutation will not form a segregated sector. The
mutant does, however, increase its relative part of the
total population in a sector of the colony (figure \ref{fig:4}).
In other words, due to the expansion of the colony, an initially tiny 
number of mutants gradually
becomes a significant part of the total population in a specific area.
Experimentally, if the mutant has some distinguishable feature --
e.g. color -- a sector will be observed.


Next we study the more interesting case of superior
mutants. In this case one observes a sector which grows faster than the
rest of the colony (see figure \ref{fig:sec346}).
In the simulations, a sharp segregation is obtained when the mutant is endowed
with a higher growth rate $\varepsilon$ (figure \ref{fig:41}) or a
higher motility (larger $D_0$, figure \ref{fig:42}). 
Figure \ref{fig:22} depicts the results obtained by simulations of the
Communicating Walker model  (with the mutant having a larger step
length, which is equivalent to a higher diffusion coefficient).
As can be seen, both models exhibit a fan-like sector of mutants, very
similar to the one observed in the experiments. The ``mixing area'',
where both strains are present, is narrow, and its width is related to
the width of the propagating front of the colony (the area where most
bacteria are alive, $b>s$).


\subsection{Branching growth}

We now turn to the case of sectoring during branching
growth.
As seen in figure \ref{fig:5}, in this case there is a slow process of segregation
even for a neutral mutation.
This results from the fact that a
particular branch may stem from a small number of bacteria, thus
allowing an initially insignificant number of mutants quickly to
become the majority in some branches, and therefore in some area of
the colony (genetic drift in small populations).

Mutants superior in motility (figure \ref{fig:6}) or growth rate
(figure \ref{fig:40}) form segregated fan-like sectors which burst out
of relatively slow advancing colony.


\subsection{The effect of chemotaxis}
As we have mentioned earlier, an additional important feature of the
bacterial movement is chemotaxis. We start by considering neutral
mutations in the case of a colony which employs repulsive chemotactic signaling.
As seen in figure \ref{fig:7}, the chemotactic response enhances the segregation of
neutral mutants. This results from 
branches being thinner in the presence of repulsive chemotaxis, and the reduced
mixing of bacteria because of the directed motion.

In the case of mutants with superior motility (figure \ref{fig:1}), a segregated sector is
formed
which is not fan-like (as opposed to the case without
chemotaxis),
probably because of the biased, radially-oriented motion of the
bacteria, coming from the long range repulsive chemotaxis.

A fan-like sector {\em does} appear when the mutant has a higher
sensitivity to the chemotactic signals (figure \ref{fig:45}). In this
case, however, the sector is composed of a mixture of the
``wild-type'' and the mutants.
Figure \ref{fig:11} displays the result of Communicating Walker
simulation for this case\footnote{The
  discrete model shows a higher tendency for segregation
  here.}.
Note the similarly of both models' results  with experimental observations (figure \ref{fig:sectors12}).

The influence of food chemotaxis on the sectors (figures
\ref{fig:66},\ref{fig:2},\ref{fig:3}) is similar to that of repulsive
chemotaxis.

Thus, the shape of the emerging sector (e.g. fan-like or not), and the
difference between the branches in the original colony and in the
sector, might testify to the nature of the advantage possessed by the mutant.
  
\section{Discussion}
\label{sec:discussion}
In this paper we presented our study of the appearance of segregated
sectors of mutants in expanding bacterial colonies. After reviewing
the experimental observations of this phenomenon, we showed the
results of simulations performed using two different models, the
discrete ``Communicating Walker'' model, and a continuous
reaction-diffusion model. Using these models as an aid to analytical
reasoning, we are able to understand what factors -- geometrical,
regulatory and others -- favor the
segregation of the mutant population. These factors include:
\begin{enumerate}
\item Expansion of the colony -- in the form of a finite front propagating
  away from areas of depleted nutrient, and towards areas of high
  nutrient concentration.
\item Branching patterns, where the population in each branch is much
  smaller than the colony's population, making a genetic drift more
  probable, so that a mutant take over the whole population in a
  sector
  of the colony. 
\item Chemotaxis: Food chemotaxis and repulsive chemotactic signaling
  cause the bacterial motion to become less random and more directed
  (outward and towards nutrients), thus lowering mixing of populations.
\item An advantageous mutant, having e.g. a higher motility or a
  faster reproduction rate, will probably conquer a sector of its own
  and quickly become segregated. This sector will usually be fan-like,
  bursting out of the colony, owing to the faster expansion of the
  mutants, as compared to that of the wild-type population.
\end{enumerate}

As in every modeling endevour, one must note the model's limitations
along with its success in reproducing and predicting biological
phenomena.
As mentioned above, care must be taken when modeling discrete entities
(i.e. bacteria) using a continuous model (for more on that, see
\cite{KL98,GKCB98,CGKB98}).
This point gains even more importance when we deal with the process
of mutation, which is a ``single bacteria event''. The discrete Walker
model is not free of this shortcoming as well, because in this model
each walker represents not one bacterium, but many \cite{BSTCCV94a}.

The observed segregation of mutant population raises some interesting
evolutionary questions. Faster movement (for example) is
an advantage for the bacteria, as the burst of sectors show. Why then
does this mutation not take over the general population and
becomes the wild-type?  
In other words, why were there any wild-type bacteria for us to isolate
in the first place?
One possible answer is that this advantage might turn out to be a 
{\em dis}advantage at different environmental conditions (e.g.
inability to remain confined to some small toxin-free oasis). Another
possibility  is that this mutant, though possessing some superior
biological feature, is lacking in another feature, essential to its
long-term survival (e.g. wasting too much energy on movement when it
is not advantageous).

Beyond the study of sectoring in bacterial colonies,
intuition about the basic mechanisms of spatial
segregation of populations might be useful for other problems.
Such problems may include the important issues of growth of
tumors and the diversification of populations on a macroscopic scale.
Both may employ similar geometrical features and communication
capabilities, leading to segregation.

Notwithstanding the above mentioned reservations, we believe this
study demonstrates once more the capability of generic models to serve
as a theoretical research tool, not only to study the basic patterns
created by bacterial colonies, but also to gain deeper understanding
of more general phenomena, such as the segregation of mutants in an
expanding colony.

\section*{Acknowledgments}
We thank J. A. Shapiro and E. Boschke for allowing us to use
photographs of their colonies.
Identifications of the
\Tname and genetic studies are carried in collaboration with the group of
D.\ Gutnick.
We are most thankful to I.\ Brains for technical assistance.
The presented studies are supported in part by a grant from
the Israeli Academy of Sciences grant no.\ 593/95 and by the
Israeli-US Binational Science Foundation BSF grant no. 00410-95.

%

\bibliographystyle{unsrt}

\begin{thebibliography}{10}

\bibitem{Darwin1859}
C.~Darwin.
\newblock {\em The Origin Of Species}.
\newblock 1859.

\bibitem{BL98}
E.~Ben-Jacob and H.~Levine.
\newblock The artistry of microorganisms.
\newblock {\em Sci. Am.}, 279(4):82--87, 1998.

\bibitem{MKNIHY92}
Matsuyama T, Kaneda K, Nakagawa Y, Isa K, H.~Hara-Hotta, and Yano I.
\newblock A novel extracellular cyclic lipopeptide which promotes
  flagellum-dependent and -independent spreading growth of {\it serratia
  marcescens}.
\newblock {\em J. Bacteriol.}, 174:1769--1776, 1992.

\bibitem{MS96}
N.~H. Mendelson and B.~Salhi.
\newblock Patterns of reporter gene expression in the phase diagram of {\bsub*}
  colony forms.
\newblock {\em J. Bacteriol.}, 178:1980--1989, 1996.

\bibitem{Kessler85}
J.~O. Kessler.
\newblock Co-operative and concentrative phenomena of swimming micro-organisms.
\newblock {\em Cont. Phys.}, 26:147--166, 1985.

\bibitem{FM89}
H.~Fujikawa and M.~Matsushita.
\newblock Fractal growth of {\bsub*} on agar plates.
\newblock {\em J. Phys. Soc. Jap.}, 58:3875--3878, 1989.

\bibitem{PK92a}
T.~J. Pedley and J.~O. Kessler.
\newblock Bioconvection.
\newblock {\em Sci. Prog.}, 76:105--123, 1989.

\bibitem{BSST92}
E.~{Ben-Jacob}, H.~Shmueli, O.~Shochet, and A.~Tenenbaum.
\newblock Adaptive self-organization during growth of bacterial colonies.
\newblock {\em Physica A}, 187:378--424, 1992.

\bibitem{MHM93}
T.~Matsuyama, R.~M. Harshey, and M.~Matsushita.
\newblock Self-similar colony morphogenesis by bacteria as the experimental
  model of fractal growth by a cell population.
\newblock {\em Fractals}, 1(3):302--311, 1993.

\bibitem{BSTCCV94a}
E.~{Ben-Jacob}, O.~Shochet, A.~Tenenbaum, I.~Cohen, A.~Czir\'ok, and T.~Vicsek.
\newblock Generic modeling of cooperative growth patterns in bacterial
  colonies.
\newblock {\em Nature}, 368:46--49, 1994.

\bibitem{BCSALT95}
E.~{Ben-Jacob}, I.~Cohen, O.~Shochet, I.~Aronson, H.~Levine, and L.~Tsimering.
\newblock Complex bacterial patterns.
\newblock {\em Nature}, 373:566--567, 1995.

\bibitem{WTMMBB95}
D.~E. Woodward, R.~Tyson, M.~R. Myerscough, J.~D. Murray, E.~O. Budrene, and
  H.~C. Berg.
\newblock Spatio-temporal patterns generated by salmonella typhimurium.
\newblock {\em Biophys. J.}, 68:2181--2189, 1995.

\bibitem{BCCVG97}
E.~{Ben-Jacob}, I.~Cohen, A.~Czir\'ok, T.~Vicsek, and D.~L. Gutnick.
\newblock Chemomodulation of cellular movement and collective formation of
  vortices by swarming bacteria and colonial development.
\newblock {\em Physica A}, 238:181--197, 1997.

\bibitem{KW97}
J.~O. Kessler and M.~F. Wojciechowski.
\newblock Collective behavior and dynamics of swimming bacteria.
\newblock In J.~A. Shapiro and M.~Dworkin, editors, {\em Bacteria as
  Mullticellular Organisms}, pages 417--450. Oxford University Press Inc., New
  York, 1997.

\bibitem{ES98}
S.~E. Esipov and J.~A. shapiro.
\newblock Kinetic model of {\em proteus mirabilis} swarm colony development.
\newblock {\em J. Math. Biol.}, 36:249--268, 1998.

\bibitem{SDA57}
M.~Doudoroff R.~Y.~Stainer and E.~A. Adelberg.
\newblock {\em The Microbial World}.
\newblock Prentice-Hall and Inc., N. J., 1957.

\bibitem{Shap88}
J.~A. Shapiro.
\newblock Bacteria as multicellular organisms.
\newblock {\em Sci. Am.}, 258(6):62--69, 1988.

\bibitem{BenJacob97}
E.~{Ben-Jacob}.
\newblock From snowflake formation to the growth of bacterial colonies. part
  {II}: Cooperative formation of complex colonial patterns.
\newblock {\em Contemp. Phys.}, 38:205--241, 1997.

\bibitem{BCL98}
E.~{Ben-Jacob}, I.~Cohen, and H.~Levine.
\newblock Cooperative self-organization of microorganisms.
\newblock {\em Adv. Phys.}, 1998.
\newblock (in press).

\bibitem{LB98}
H.~Levine and E.~{Ben-Jacob}.
\newblock The art and science of microorganisms.
\newblock {\em Sci. Am.}, 279(4), 1998.

\bibitem{Mend78}
N.~H. Mendelson.
\newblock Helical {{\bsub*}} macrofibers: morphogenesis of a bacterial
  multicellular macroorganism.
\newblock {\em Proc. Natl. Acad. Sci. USA}, 75(5):2478--2482, 1978.

\bibitem{Devreotes89}
P.~Devreotes.
\newblock {\it Dictyostelium discoideum}: a model system for cell-cell
  interactions in development.
\newblock {\em Science}, 245:1054--1058, 1989.

\bibitem{Harshey94}
R.~M. Harshey.
\newblock Bees aren't the only ones: swarming in gram-negative bacteria.
\newblock {\em Molecular Microbiology}, 13:389--394, 1994.

\bibitem{FWG94}
Fuqua WC, Winans SC, and Greenberg EP.
\newblock Quorum sensing in bacteria: the {LuxR-LuxI} family of cell
  density-responsive transcriptional regulators.
\newblock {\em J. Bacteriol.}, 176:269--275, 1994.

\bibitem{LWFBSLW95}
A.~Latifi, M.~K. Winson, M.~Foglino, B.~W. Bycroft, G.~S. Stewart,
  A.~Lazdunski, and P.~Williams.
\newblock Multiple homologues of {LuxR} and {LuxI} control expression of
  virulence determinants and secondary metabolites through quorum sensing in
  {{\it Pseudomonas aeruginosa}} {PAO1}.
\newblock {\em Molecular Microbiology}, 17:333--343, 1995.

\bibitem{FWG96}
C.~Fuqua, S.~C. Winans, and E.~P. Greenberg.
\newblock Census and consensus in bacterial ecosystems: the {LuxR-LuxI} family
  of quorum-sensing transcriptional regulators.
\newblock {\em Annu. Rev. Microbiol.}, 50:727--751, 1996.

\bibitem{BB91}
E.~O. Budrene and H.~C. Berg.
\newblock Complex patterns formed by motile cells of {\em esherichia coli}.
\newblock {\em Nature}, 349:630--633, 1991.

\bibitem{BE95}
Y.~Blat and M.~Eisenbach.
\newblock Tar-dependent and -independent pattern formation by {\salmon*}.
\newblock {\em J. Bac.}, 177(7):1683--1691, 1995.

\bibitem{BB95}
E.~O. Budrene and H.~C. Berg.
\newblock Dynamics of formation of symmetrical patterns by chemotactic
  bacteria.
\newblock {\em Nature}, 376:49--53, 1995.

\bibitem{ST91}
J.~A. Shapiro and D.~Trubatch.
\newblock Sequential events in bacterial colony morphogenesis.
\newblock {\em Physica D}, 49:214--223, 1991.

\bibitem{SM93}
B.~Salhi and N.~H. Mendelson.
\newblock Patterns of gene expression in {\bsub*} colonies.
\newblock {\em J. Bacteriol.}, 175:5000--5008, 1993.

\bibitem{GR95}
T.~Galitski and J.~R. Roth.
\newblock Evidence that {F} plasmid transfer replication underlies apparent
  adaptive mutation.
\newblock {\em Science}, 268:421--423, 1995.

\bibitem{RPF95}
J.~P. Rasicella, P.~U. Park, and M.~S. Fox.
\newblock Adaptive mutation in {\ecoli*} : a role for conjugation.
\newblock {\em Science}, 268:418--420, 1995.

\bibitem{Miller98}
R.~V. Miller.
\newblock Bacterial gene swapping in nature.
\newblock {\em Sci. Am.}, 278(1), 1998.

\bibitem{MF90}
M.~Matsushita and H.~Fujikawa.
\newblock Diffusion-limited growth in bacterial colony formation.
\newblock {\em Physica A}, 168:498--506, 1990.

\bibitem{FM91}
H.~Fujikawa and M.~Matsushita.
\newblock Bacterial fractal growth in the concentration field of nutrient.
\newblock {\em J. Phys. Soc. Jap.}, 60:88--94, 1991.

\bibitem{SC38}
R.~N. Smith and F.~E. Clacrk.
\newblock Motile colonies of bacillus alvei and other bacteria.
\newblock {\em J. Bacteriol.}, 35:59--60, 1938.

\bibitem{Henrici48}
T.~H. Henrici.
\newblock {\em The Biology of Bacteria: The Bacillaceae}.
\newblock D. C. Heath \& company, 3rd edition, 1948.

\bibitem{WitSan81}
T.~A. Witten and L.~M. Sander.
\newblock Diffusion--limited aggregation, a kinetic critical phenomenon.
\newblock {\em Phys. Rev. Lett.}, 47:1400, 1981.

\bibitem{Sander86}
L.M. Sander.
\newblock Fractal growth processes.
\newblock {\em Nature}, 322:789--793, 1986.

\bibitem{Vicsek89}
T.~Vicsek.
\newblock {\em Fractal Growth Phenomena}.
\newblock World Scientific, New York, 1989.

\bibitem{MM95}
T.~Matsuyama and M.~Matsushita.
\newblock Morphogenesis by bacterial cells.
\newblock In P.~M. Iannaccone and M.~K. Khokha, editors, {\em Farctal Geometry
  in Biological Systems, an Analytical Approach}, pages 127--171. CRC Press,
  New-York, 1995.

\bibitem{BTSA94}
E.~{Ben-Jacob}, A.~Tenenbaum, O.~Shochet, and O.~Avidan.
\newblock Holotransformations of bacterial colonies and genome cybernetics.
\newblock {\em Physica A}, 202:1--47, 1994.

\bibitem{TBG98}
M.~Tcherpikov, E.~{Ben-Jacob}, and D.~Gutnick.
\newblock Identification of two pattern-forming strains and their localization
  in a phylogenetic cluster.
\newblock {\em Int. J. Syst. Bacteriol.}, 1998.
\newblock (in press).

\bibitem{Shapiro95b}
J.~A. Shapiro.
\newblock The significances of bacterial colony patterns.
\newblock {\em BioEssays}, 17(7):597--607, 1995.

\bibitem{BB98}
E.~Boschke and Th. Bley.
\newblock Growth patterns of yeast colonies depending on nutrient supply.
\newblock {\em Acta Biotechnol.}, 18(1):17--27, 1998.

\bibitem{Fisher37}
R.~A. Fisher.
\newblock The wave of advance of advantageous genes.
\newblock {\em Annual Eugenics}, 7:255--369, 1937.

\bibitem{GKCB98}
I.~Golding, Y.~Kozlovsky, I.~Cohen, and E.~{Ben-Jacob}.
\newblock Studies of bacterial branching growth using reaction-diffusion models
  of colonial development.
\newblock {\em Physica A}, 260(3-4):510--554, 1998.

\bibitem{PS78}
H.~Parnas and L.~Segel.
\newblock A computer simulation of pulsatile aggregation in {{\it Dictyostelium
  discoideum}}.
\newblock {\em J. Theor. Biol.}, 71:185--207, 1978.

\bibitem{Mackay78}
S.~A. Mackay.
\newblock Computer simulation of aggregation in dictyostelium discoideum.
\newblock {\em J. Cell. Sci.}, 33:1--16, 1978.

\bibitem{BCSCV95}
E.~{Ben-Jacob}, I.~Cohen, O.~Shochet, A.~Czir\'ok, and T.~Vicsek.
\newblock Cooperative formation of chiral patterns during growth of bacterial
  colonies.
\newblock {\em Phys. Rev. Lett.}, 75(15):2899--2902, 1995.

\bibitem{KL93}
D.~A. Kessler and H.~Levine.
\newblock Pattern formation in {\em dictyostelium} via the dynamics of
  cooperative biological entities.
\newblock {\em Phys. Rev. E}, 48:4801--4804, 1993.

\bibitem{KLT97}
D.~A. Kessler, H.~Levine, and L.~Tsimring.
\newblock Computational modeling of mound development in {\em dictyostelium}.
\newblock {\em Physica D}, 106(3-4):375--388, 1997.

\bibitem{Mimura97}
M.~Mimura, H.~Sakaguchi, and M.~Matsushita.
\newblock A reaction-diffusion approach to bacterial colony formation.
\newblock {\em preprint}, 1997.

\bibitem{MWIRMSM98}
M.~Matsushita, J.~Wakita, H.~Itoh, I.~Rafols, T.~Matsuyama, H.~Sakaguchi, and
  M.~Mimura.
\newblock Interface growth and pattern formation in bacterial colonies.
\newblock {\em Physica A}, 249:517--524, 1998.

\bibitem{KMMUS97}
K.~Kawasaki, A.~Mochizuki, M.~Matsushita, T.~Umeda, and N.~Shigesada.
\newblock Modeling spatio-temporal patterns created by bacillus-subtilis.
\newblock {\em J. Theor. Biol.}, 188:177--185, 1997.

\bibitem{Kitsunezaki97}
S.~Kitsunezaki.
\newblock Interface dynamics for bacterial colony formation.
\newblock {\em J. Phys. Soc. Jpn}, 66(5):1544--1550, 1997.

\bibitem{Rafols98}
I. Rafols, {\em Formation of concentric rings in bacterial colonies}, MSc
  thesis, Chuo University, Japan, 1998.

\bibitem{CGKB98}
I.~Cohen, I.~Golding, Y.~Kozlovsky, and E.~{Ben-Jacob}.
\newblock Continuous and discrete models of cooperation in complex bacterial
  colonies.
\newblock In E.~Inan and K.~Z. Markov, editors, {\em Continuum Models and
  Discrete Systems: Proceedings of the 9th International Symposium}. World
  Scientific Publishing, 1998.

\bibitem{KL98}
D.~A. Kessler and H.~Levine.
\newblock Fluctuation-induced diffusive instabilities.
\newblock {\em Nature}, 394(6):556--558, 1998.

\bibitem{Darwin1859a}
C.~Darwin.
\newblock {\em The Origin Of Species}.
\newblock 1859.
\newblock "I have hitherto sometimes spoken as if the variations ... were due
  to chance. This, of course, is a wholly incorrect expression, but it serves
  to acknowledge plainly our ignorance of the cause of each particular
  variation. ... [The facts] lead to the conclusion that variability is
  generally related to the conditions of life to which each species has been
  exposed during several successive generations." Chap. V.

\bibitem{COM88}
J.~Cairns, J.~Overbaugh, and S.~Miller.
\newblock The origin of mutants.
\newblock {\em Nature}, 335:142--145, 1988.

\bibitem{Hall88}
B.~G. Hall.
\newblock Adaptive evolution that requires multiple spontaneous mutations. {I}.
  mutations involving an insertion sequence.
\newblock {\em Genetics}, 120:887--897, 1988.

\bibitem{Hall91}
B.~G. Hall.
\newblock Adaptive evolution that requires multiple spontaneous mutations:
  mutations involving base substitutions.
\newblock {\em Proc. Natl. Acad. Sci. USA}, 88:5882--5886, 1991.

\bibitem{Foster93}
P.~L. Foster.
\newblock Adaptive mutation: The uses of adversity.
\newblock {\em Annu. Rev. Microbiol.}, 47:467--504, 1993.

\bibitem{BSTCCV94b}
E.~{Ben-Jacob}, O.~Shochet, A.~Tenenbaum, I.~Cohen, A.~Czir\'ok, and T.~Vicsek.
\newblock Communication, regulation and control during complex patterning of
  bacterial colonies.
\newblock {\em Fractals}, 2(1):15--44, 1994.

\bibitem{CCB96}
I.~Cohen, A.~Czir\'ok, and E.~{Ben-Jacob}.
\newblock Chemotactic-based adaptive self organization during colonial
  development.
\newblock {\em Physica A}, 233:678--698, 1996.

\bibitem{BSCTCV95}
E.~Ben-Jacob, O.~Shochet, I.~Cohen, A.~Tenenbaum, A.~Czir\'ok, and T.~Vicsek.
\newblock Cooperative strategies in formation of complex bacterial patterns.
\newblock {\em Fractals}, 3:849--868, 1995.

\bibitem{Adler69}
J.~Adler.
\newblock Chemoreceptors in bacteria.
\newblock {\em Science}, 166:1588--1597, 1969.

\bibitem{BP77}
H.~C. Berg and E.~M. Purcell.
\newblock Physics of chemoreception.
\newblock {\em Biophysical Journal}, 20:193--219, 1977.

\bibitem{Lacki81}
J.~M. Lackiie, editor.
\newblock {\em Biology of the chemotactic response}.
\newblock Cambridge Univ. Press, 1986.

\bibitem{Berg93}
H.~C. Berg.
\newblock {\em Random Walks in Biology}.
\newblock Princeton University Press, Princeton, N.J., 1993.

\bibitem{KCGB98}
Yonathan Kozlovsky, Inon Cohen, Ido Golding and Eshel Ben-Jacob, submitted to
  Phys. Rev. E.

\bibitem{Murray89}
J.~D. Murray.
\newblock {\em Mathematical Biology}.
\newblock Springer-Verlag, Berlin, 1989.

\bibitem{SD96}
J.~A. Shapiro and M.~Dworkin, editors.
\newblock {\em Bacteria as Multicellular Organisms}.
\newblock Oxford University Press, New-York, 1997.

\end{thebibliography}


\begin{figure}
\caption {
\label{fig:physa1}
Typical example of branching growth of the \Tname for $1g/l$
peptone level and 1.5\% agar concentration.
}
\end{figure}

\begin{figure}
\caption {
\label{fig:shapiro}
Emerging sector in a {\em E. Coli} colony. Picture by James A.
Shapiro, 
From ``Bacteria as Multicellular Organisms'' edited by J. Shapiro and M.
Dworkin  \protect{\cite{SD96}}. (C) Copyright 1997 by Oxford
University Press, Inc.  
 Used by
permission of Oxford University Press. 
}

\end{figure}

\begin{figure}
\caption {
\label{fig:boschke}
Emerging sector in a colony of {\em Yarrowia lipolytica}. Taken from
\protect{\cite{BB98}}, used with permission.
}
\end{figure}

\begin{figure}
\caption {
\label{fig:physa2}
Examples of typical  patterns of \Tname for intermediate agar
concentration.
(top left) At very high peptone level (peptone 12$g/l$, agar concentration 1.75\%)
the pattern is compact.
(top right) At high peptone level (3$g/l$, agar 2\%) the pattern is of dense
fingers with pronounced radial symmetry -- similar to patterns
observed in Hele-Show cell.
(bottom left) At intermediate peptone level (1$g/l$, agar 1.75\%) the pattern is
"bushy" fractal-like pattern, with branch width smaller than the
distance between branches.
(bottom right) At low peptone level (0.1$g/l$, agar 1.75\%) there are fine radial
branches with apparent circular envelope.
}
\end{figure}

\begin{figure}
\caption {
\label{fig:finger}
A closer look on branches of a colony:
(left) Numarsky (polarized light) microscopy shows the hight of the
branches and their envelope. What is actually seen is the layer of
lubrication fluid, not the bacteria.
(right) X50 magnification shows the bacteria inside a branch. Each bar
is a single bacterium. There are no bacteria outside the branch.
}
\end{figure}

\begin{figure}
\caption {
\label{fig:sec11}
A compact growth pattern of \Tname, obtained when the agar surface is
very soft ($0.4 \%$ agar concentration and 0.1 g/l peptone).
}
\end{figure}

\begin{figure}
\caption {
\label{fig:sec346}
Emerging sectors in compact colonies of {\it P. dendritiformis}.
 (left) \Tvar, 10 g/l peptone, $0.5 \%$ agar.
 (middle) \Cvar, 1.5 g/l peptone, $0.4 \%$ agar.
 (right) \Cvar, 10 g/l peptone, $0.4 \%$ agar.
}
\end{figure}

\begin{figure}
\caption {
\label{fig:sectors12}
Emerging sectors in branching colonies of \Tname, obtained at 1 g/l
peptone and $1.75 \%$ agar (left), 0.8 g/l peptone and $2 \%$ agar (right). 
}
\end{figure}

\begin{figure}
\caption {
\label{fig:sec57}
Emerging sectors in branching colonies of \Tname, obtained at 5 g/l
peptone and $1.75 \%$ agar, in the presence of the antibiotics Stromocine.
}
\end{figure}


\begin{figure}
\caption {
\label{fig:4}
Neutral mutation in a compact colony: Results of simulation of the
continuous model.
It is seen that the wildtype covers the complete colony area uniformly
(left). However, the mutants-to-wildtype ratio increases in a sector
of the colony (right), indicating the tendency toward segregation.
}
\end{figure}

\begin{figure}
\caption {
\label{fig:41}
Mutant with a higher growth rate ($\varepsilon_2>\varepsilon_1$ in the
continuous model), compact colony.
(left) The mutant (light) irrupts in a fan-like sector from the
wildtype (dark) colony.
(right) The
wildtype does not penetrate the mutation sector.
}
\end{figure}

\begin{figure}
\caption {
\label{fig:42}
Mutant with a higher motility ($D_{0_2}>D_{0_1}$ in the
continuous model), compact colony.
(left) The mutant (light) irrupts in a slice-like sector from the
wildtype (dark) colony. 
(right) The
wildtype does not penetrate the mutation sector (right).
}
\end{figure}

\begin{figure}
\caption {
\label{fig:22}
Mutant with a higher motility (larger step length in the
Communicating Walker model), compact colony.
}
\end{figure}

\begin{figure}
\caption {
\label{fig:5}
Neutral mutation in a branching colony: Results of simulation of the
continuous model.
It is seen that the mutant gradually becomes a majority of the
population in a sector of the colony (right), while the wildtype is
gradually ``expelled'' from this area (left).
}
\end{figure}

\begin{figure}
\caption {
\label{fig:6}
Mutant with a higher motility ($D_{0_2}>D_{0_1}$ in the
continuous model), branching colony.
(left) The mutant (light) irrupts in a sector from the
wildtype (dark) colony.
(right) The
wildtype does not penetrate the mutation sector.
}
\end{figure}

\begin{figure}
\caption {
\label{fig:40}
Mutant with a higher growth rate ($\varepsilon_2>\varepsilon_1$ in the
continuous model), branching colony.
(left) The mutant (light) irrupts in a fan-like sector from the
wildtype (dark) colony.
(right) The
wildtype does not penetrate the mutation sector.
}
\end{figure}

\begin{figure}
\caption {
\label{fig:7}
Neutral mutation in a branching colony, with the presence of repulsive
chemotactic signaling: Results of simulation of the
continuous model.
It is seen that the mutant gradually becomes a majority of the
population in a sector of the colony (right), while the wildtype is
gradually ``expelled'' from this area (left).
}
\end{figure}

\begin{figure}
\caption {
\label{fig:1}
Mutant with a higher motility ($D_{0_2}>D_{0_1}$ in the
continuous model), branching colony with presence of repulsive
chemotactic signaling.
The mutant (light) irrupts in a sector from the
wildtype (dark) colony (left). 
Note that the sector does not burst out of the rest of the colony.
(right) The
wildtype does not penetrate the mutation sector.
}
\end{figure}

\begin{figure}
\caption {
\label{fig:45}
Mutant with a higher sensitivity to repulsive chemotactic signaling ($\chi_{0_2}>\chi_{0_1}$ in the
continuous model), branching colony.
The mutant irrupts in a fan-like sector from the
colony of wildtype bacteria (left).
However, the sector is not a segregated area, and contains wildtype
bacteria as well (right).
}
\end{figure}

\begin{figure}
\caption {
\label{fig:11}
Mutant with a higher sensitivity to repulsive chemotactic signaling:
Results of the Communicating Walker
model, branching colony. as in the continuous model, 
the mutant irrupts in a fan-like sector from the
 colony of wildtype bacteria.
}
\end{figure}

\begin{figure}
\caption {
\label{fig:66}
Neutral mutation in a branching colony, with the presence of food
chemotaxis: Results of simulation of the
continuous model.
It is seen that the mutant gradually becomes a majority of the
population in a sector of the colony (right), while the wildtype is
gradually ``expelled'' from this area (left).
}
\end{figure}

\begin{figure}
\caption {
\label{fig:2}
Mutant with a higher sensitivity to food chemotaxis ($\chi_{0_2}>\chi_{0_1}$ in the
continuous model), branching colony.
(left) The mutant (light) irrupts in a fan-like sector from the
wildtype (dark) colony.
(right) The
wildtype does not penetrate the mutation sector.
}
\end{figure}

\begin{figure}
\caption {
\label{fig:3}
Mutant with a higher motility ($D_{0_2}>D_{0_1}$ in the
continuous model), branching colony with presence of food chemotaxis.
(left) The mutant (light) irrupts in a sector from the
wildtype (dark) colony. 
Note that the sector does not burst out of the rest of the colony.
(right) The
wildtype does not penetrate the mutation sector.
}
\end{figure}

\end{document}